\documentclass[10pt,twocolumn,english,jcp, aip]{revtex4-1}
\usepackage[T1]{fontenc}
\usepackage[latin9]{inputenc}
\usepackage{amsmath}
\usepackage{amssymb}
\usepackage{graphicx}
\usepackage{esint}

\makeatletter

\providecommand{\tabularnewline}{\\}

\@ifundefined{date}{}{\date{}}
\usepackage{babel}
\usepackage{enumitem}

\makeatother

\usepackage{babel}
\begin{document}

\title{Overlapped Embedded Fragment Stochastic Density Functional Theory
for Covalently Bonded Materials}

\title{Overlapped Embedded Fragment Stochastic Density Functional Theory
for Covalently Bonded Materials}

\author{Ming Chen}

\affiliation{Department of Chemistry, University of California, and Lawrence Berkeley
National Laboratory, Berkeley, California 94720, USA}

\author{Roi Baer}

\affiliation{Fritz Haber Center of Molecular Dynamics and Institute of Chemistry,
The Hebrew University of Jerusalem, Jerusalem, 91904 Israel}

\author{Daniel Neuhauser}

\affiliation{Department of Chemistry and Biochemistry, University of California,
Los Angeles, California 90095, USA}

\author{Eran Rabani}

\affiliation{Department of Chemistry, University of California, and Lawrence Berkeley
National Laboratory, Berkeley, California 94720, USA}

\affiliation{The Raymond and Beverly Sackler Center of Computational Molecular
and Materials Science, Tel Aviv University, Tel Aviv 69978, Israel}
\begin{abstract}
The stochastic density functional theory (DFT) {[}Phys. Rev. Lett.
111, 106402 (2013){]} is a valuable linear scaling approach to Kohn-Sham
DFT that does not rely on the sparsity of the density matrix. Linear
(and often sub-linear) scaling is achieved by introducing a controlled
statistical error in the density, energy and forces. The statistical
error (noise) is proportional to the inverse square root of the number
of stochastic orbitals and thus decreases slowly, however, by dividing
the system to fragments that are embedded stochastically, the statistical
error can be reduced significantly. This has been shown to provide
remarkable results for non-covalently bonded systems, however, the
application to covalently bonded systems had limited success, particularly
for delocalized electrons. Here, we show that the statistical error
in the density correlates with both the density and the density matrix
of the system and propose a new fragmentation scheme that elegantly
interpolates between overlapped fragments. We assess the performance
of the approach for bulk silicon of varying supercell sizes (up to
$N_{e}=16384$ electrons) and show that overlapped fragments reduce
significantly the statistical noise even for systems with a delocalized
density matrix.
\end{abstract}
\maketitle

\section{Introduction}

An accurate description of the electronic properties is often a prerequisite
for understand the behavior of complex materials. Density functional
theory~\citep{PhysRev.136.B864} within the Kohn-Sham (KS) formulation~\citep{PhysRev.140.A1133}
provides an excellent framework that balances computational complexity
and accuracy and thus, has been the method of choice for extended,
large-scale systems. Solving the KS equations scales formally as $O\left(N_{e}^{3}\right)$
where $N_{e}$ is the number of electron in the system. Traditional
implementations of KS-DFT are often limited to relatively systems
containing $N_{e}<10^{4}$ electrons,\citep{RevModPhys.73.515,R0953-8984-14-11-313,RevModPhys.86.253}
even with the high-performance computer architectures of today. Clearly,
there is a need for low-scaling KS-DFT approaches. 

A natural approach to linear-scaling KS-DFT relies on the ``locality''
of the density matrix, that is, elements of the density matrix $\rho(\boldsymbol{r},\boldsymbol{r}')$
decay with increasing $\|\boldsymbol{r}-\boldsymbol{r}'\|$.\citep{PhysRevB.47.9973,PhysRevB.48.14646,GOEDECKER1995261,PhysRevB.51.10157,PhysRevLett.76.3168,PhysRevLett.79.3962,PhysRevB.58.12704,RevModPhys.71.1085,doi:10.1063/1.1590632}
The reliance on Kohn\textquoteright s \textquoteleft \textquoteleft nearsightedness\textquoteright \textquoteright{}
principle~\citep{PhysRevLett.76.3168} makes these approaches sensitive
to the dimensionality and the character of the system. They work extremely
well for low dimensional structures or systems with a large fundamental
gap,\citep{PhysRevLett.79.3962} but in 3-dimensions (3D), linear
scaling is achieved only for very large systems, typically for $N_{e}>10^{4}$.
An alternative to Kohn\textquoteright s \textquoteleft \textquoteleft nearsightedness\textquoteright \textquoteright{}
principle is based on ``divide'' and ``conquer'',\citep{PhysRevLett.66.1438,PhysRevB.44.8454,PhysRevB.53.12713,PhysRevB.77.085103}
where the density or the density matrix is partitioned into fragments
and the KS equations are solved for each fragment separately using
localized basis sets.\citep{PhysRevLett.66.1438,PhysRevB.53.12713,PhysRevB.77.085103}
The complexity of solving the KS equations is shifted to computing
the interactions between the fragments,\citep{PhysRevB.53.12713,doi:10.1021/ct9000119,doi:10.1021/ct9001784,doi:10.1063/1.3659293,doi:10.1063/1.3582913,YuE10861}
and the accuracy (systematic errors that decrease with the fragment
size) and scaling depend on the specific implementation. 

Recently, we have proposed a new scheme for linear scaling DFT based
approach.\citep{PhysRevLett.111.106402} Stochastic DFT (sDFT) sidesteps
the calculation of the density matrix and is therefore, not directly
sensitive to its evasive sparseness. Instead, the density is given
in terms of a trace formula and is evaluated using stochastic occupied
orbitals generated by a Chebyshev or Newton expansion of the occupation
operator.\citep{PhysRevLett.111.106402} Random fluctuations of local
properties, e.g., energy per atom, forces, and density of states,
are controlled by the number of stochastic orbitals and are often
independent of the system size,\citep{Roisubmit} leading to a computational
cost that scales linearly (and sometimes sub-linearly). 

The statistical error is substantially reduced when, instead of sampling
the full density, one samples the difference between the full system
density and that of reference fragments.\citep{doi:10.1063/1.4890651}
Indeed, embedded fragment stochastic DFT (efsDFT) has been extremely
successful in reducing the noise level for non-covalently bonded fragment.\citep{doi:10.1063/1.4890651}
However, for covalently bonded fragments, efsDFT needs to be structured
appropriately,\citep{doi:10.1063/1.4984931} particularly for periodic
boundary conditions.

In this manuscript, we have developed a new fragmentation approach
suitable for covalently bonded systems with open or periodic boundary
conditions (PBC). Similar to efsDFT, we divide the system into \emph{core}
fragments, but in addition, for each core fragment we add a \emph{buffer}
zone allowing for fragment to overlap in the embedding procedure.
Similar in spirit to earlier work on divide and conquer,\citep{PhysRevB.53.12713}
we demonstrate that allowing the fragments to overlap results in significant
improvements of the reference density matrix, reducing significantly
the level of statistical noise. The manuscript is organized as following:
In Sec.~\ref{sec:Stochastic-Density-Functional} we briefly review
the theory of stochastic orbital DFT and its embedded-fragmented version.
In Sec.~\ref{sec:Overlapped-Embedded-Fragmented} we outline the
new overlapped embedded fragmented stochastic DFT (o-efsDFT). Sec.~\ref{sec:Results-and-Discussion}
presents the results of numerical calculations on a set of silicon
crystals of varying super-cell size, up to $4096$ atoms and discusses
the significance of the overlapped fragments on reducing the statistical
error.

\section{Stochastic Density Functional Theory and Embedded Fragment Stochastic
Density Functional Theory\label{sec:Stochastic-Density-Functional}}

We consider a system with periodic boundary condition in a super-cell
of volume $V$ with an electron density $\rho(\boldsymbol{r})$ represented
on a real-space grid with $N_{G}$ grid points. The electronic properties
are described within Kohn-Sham (KS) density functional theory,\citep{PhysRev.140.A1133}
where the density is given by $\rho\left(\boldsymbol{r}\right)=\sum_{i=1}^{N_{\text{occ}}}\langle\boldsymbol{r}|\psi_{i}\rangle\langle\psi_{i}|\boldsymbol{r}\rangle$.
Here, $\langle\boldsymbol{r}|\psi_{i}\rangle$ are the KS orbitals
and $N_{\text{occ}}$ is the number of occupied states. The KS density
can also be expressed as a trace over the density matrix, $\hat{\rho}=\lim_{\beta\rightarrow\infty}\theta_{\beta}(\mu-\hat{h}_{\text{KS}})$:
\begin{equation}
\rho\left(\boldsymbol{r}\right)=\text{Tr}\,\theta_{\beta}(\mu-\hat{h}_{\text{KS}})\delta\left(\boldsymbol{r}-\hat{\boldsymbol{r}}\right),\label{eq:rho}
\end{equation}
where $\delta\left(\boldsymbol{r}-\hat{\boldsymbol{r}}\right)$ is
Dirac's delta function and $\theta_{\beta}\left(\hat{x}\right)=\left(1+e^{-\beta\hat{x}}\right)^{-1}$
is a step function in the limit $\beta\rightarrow\infty$ (in practice
$\beta$ is chosen to be large enough to converge the ground state
properties). The chemical potential, $\mu$, is determined by imposing
the relation $N_{e}\left(\mu\right)=\text{Tr}\theta_{\beta}(\mu-\hat{h}_{\text{KS}})$
where $N_{e}\left(\mu\right)$ is the total number of electrons in
the system. In Eq.~\eqref{eq:rho}, $\hat{h}_{KS}$ is the KS Hamiltonian,
given by:
\begin{equation}
\hat{h}_{\text{KS}}=\hat{t}+\hat{v}_{\text{loc}}+\hat{v}_{\text{nl}}+\hat{v}_{\text{H}}+\hat{v}_{\text{XC}},\label{eq:KS Hamiltonian}
\end{equation}
where $\hat{t}$ is the kinetic energy, $\hat{v}_{\text{loc}}$ and
$\hat{v}_{\text{nl}}$ are the local and nonlocal pseudopotentials,
$\hat{v}_{\text{H}}$ is the Hartree potential, and $\hat{v}_{\text{XC}}$
is the exchange-correlation potential. 

Eq.~\eqref{eq:rho} is the starting point for the derivation of the
stochastic DFT.\citep{PhysRevLett.111.106402} The trace is performed
using $N_{\chi}$ stochastic orbitals, $|\chi\rangle$, leading to:
\begin{align}
\rho\left(\boldsymbol{r}\right) & =\left\langle \left\langle \chi\left|\theta_{\beta}(\mu-\hat{h}_{\text{KS}})\delta\left(\boldsymbol{r}-\hat{\boldsymbol{r}}\right)\right|\chi\right\rangle \right\rangle _{\chi}\nonumber \\
= & \left\langle \left\langle \xi\left|\delta\left(\boldsymbol{r}-\hat{\boldsymbol{r}}\right)\right|\xi\right\rangle \right\rangle _{\chi}\equiv\left\langle \left|\xi\left(\boldsymbol{r}\right)\right|^{2}\right\rangle _{\chi}\,\,,\label{eq:rho-stoch}
\end{align}
where $|\xi\rangle=\sqrt{\theta_{\beta}(\mu-\hat{h}_{\text{KS}})}|\chi\rangle$
is a projected stochastic orbital. In the above, $\chi\left(\boldsymbol{r}\right)=\pm\left(\Delta V\right)^{-1/2}$,
where $\Delta V=V/N_{G}$ is the volume element of the real-space
grid and $\pm$ indicates that we randomly, uniformly, and independently,
select the sign of $\chi$ for each grid point $\boldsymbol{r}.$
The additional brackets ($\left\langle \cdots\right\rangle _{\chi}$)
in Eq.~\eqref{eq:rho-stoch} represent an average over all stochastic
realizations, namely, $\left\langle \cdots\right\rangle _{\chi}=\frac{1}{N_{\chi}}\sum_{\chi}\cdots$.
Eq.~\eqref{eq:rho-stoch} becomes exact only in the limit that $N_{\chi}\rightarrow\infty$
and otherwise it is a stochastic approximation to the electron density,
with random error whose magnitude decays as $\sqrt{N_{\chi}}$. Eq.~\eqref{eq:rho-stoch}
is solved self-consistently, since the KS Hamiltonian depends on $\rho\left(\boldsymbol{r}\right)$
itself. 

Obtaining the total energy within the sDFT formalism is tricky. The
Hartree, local pseudopotential, and local exchange-correlation energy
terms can be obtained directly from the stochastic density, similar
to deterministic DFT. The calculation of the kinetic energy and nonlocal
pseudopotential energy requires the Kohn-Sham orbitals in deterministic
DFT. In sDFT, these can be evaluated using the relations:\citep{PhysRevLett.111.106402}
\begin{align}
E_{K} & =\text{Tr}\,\hat{\rho}\,\hat{t}=\left\langle \left\langle \xi\left|\hat{t}\right|\xi\right\rangle \right\rangle _{\chi}\\
E_{\text{nl}} & =\text{Tr}\,\hat{\rho}\,\hat{v}_{nl}=\left\langle \left\langle \xi\left|\hat{v}_{\text{nl}}\right|\xi\right\rangle \right\rangle _{\chi}.
\end{align}
The advantage of the stochastic DFT formalism is that the density
can be calculated with linear (and often sub-linear) scaling at the
cost of introducing a controlled statistical error, which is controlled
by the number of stochastic orbitals as and decreases with $\sqrt{N_{\chi}}$.
Linear scaling is achieved by taking advantage of the sparsity of
$\hat{h}_{\text{KS}}$ in real- and reciprocal-space representations
and approximating $\xi\left(\boldsymbol{r}\right)=\sqrt{\theta_{\beta}(\mu-\hat{h}_{\text{KS}})}\,\chi\left(\boldsymbol{r}\right)$
in Eq.~\eqref{eq:rho-stoch} using a Chebyshev or Newton interpolation
polynomials,\citep{Kosloff1988,Kosloff1994} cf. $\sqrt{\theta_{\beta}(\mu-\hat{h}_{\text{KS}})}\approx\sum_{i=0}^{N_{c}}a_{i}(\mu)T_{n}(\hat{h}_{\text{KS}})$
where $N_{c}$ is the length of polynomial chain and $T_{n}\text{'s}$
are the Chebyshev polynomials. In situations where the statistical
noise does not increase with the system size, sDFT scales linearly
with the system size. Often, for certain local observables, we find
that sDFT scales better than linear (sub-linear) due to self-averaging.
This has been recently discussed in detail in Ref.~\onlinecite{Roisubmit}.

\begin{figure}[t]
\centering \includegraphics[width=8cm]{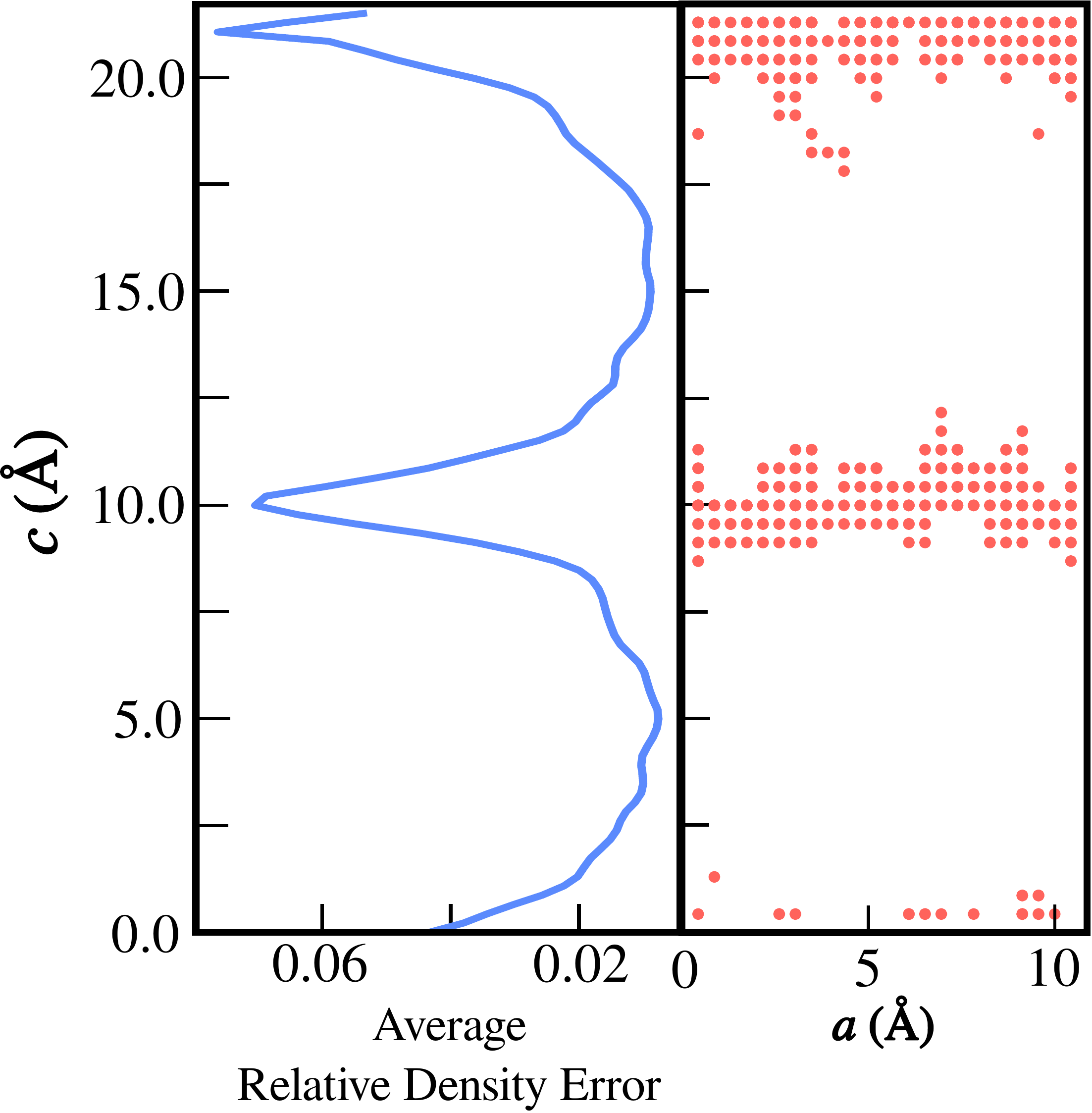} \caption{The average relative deviation of the stochastic density from the
deterministic density along the c-axis of a silicon supper cell with
$128$ atoms ($2\times2\times4$) (left panel). The stochastic density
was calculated within efsDFT with two non-overlapped $\text{Si}_{64}$
fragments ($2\times2\times2$). The right panel shows this relative
density difference in projected onto 2D (we only plot deviation above
$10\%$). The largest deviations is at the interface between the two
fragments ($c\approx0,10,20\text{Å}$). }
\label{fig:bad-rho} 
\end{figure}

The decay of the magnitude of the statistical error with $\sqrt{N_{\chi}}$
implies that in order to reduce the error by an \emph{order} of magnitude,
one has to increase the number of stochastic orbitals by \emph{two}
\emph{orders} of magnitude. Thus, small statistical errors required
to converge local and single particle observables, often require a
huge set of stochastic orbitals, making sDFT impractical. Since the
standard deviation of the density is proportional to $\rho$$\left(\boldsymbol{r}\right)$
itself (as shown in the supplementary information), the statistical
noise can be reduced by decomposing the density into a reference density,
$\rho_{\text{ref }}\left(\boldsymbol{r}\right)$, and a correction
terms $\Delta\rho\left(\boldsymbol{r}\right)\ll\rho\left(\boldsymbol{r}\right)$,
such that $\rho\left(\boldsymbol{r}\right)=\rho_{\text{ref}}\left(\boldsymbol{r}\right)+\Delta\rho\left(\boldsymbol{r}\right)$.
This is the central idea behind embedded-fragmented stochastic DFT
(efsDFT).\citep{doi:10.1063/1.4890651,doi:10.1063/1.4984931,Roisubmit}
The system is divided into non-overlapping fragments and the reference
density is decomposed as a sum of the fragment densities. This has
led to a significant reduction of the noise in efsDFT compared to
sDFT when fragments are closed-shell molecules in non-covalently bonded
systems.\citep{doi:10.1063/1.4890651} However, for covalently bonded
materials it is necessary to break covalent bonds when fragmenting
the system. This leads to an increasingly large statistical error
of the density at the boundaries between the fragments (see Fig.~\ref{fig:bad-rho}
for an illustration of this effect for $\text{Si}_{128}$ with PBCs
where the system was divided into two non-overlapping $\text{Si}_{64}$
fragments), leading to small improvements of efsDFT with respect to
sDFT.\citep{doi:10.1063/1.4984931,Roisubmit} 

\section{Overlapped Embedded Fragmented stochastic DFT\label{sec:Overlapped-Embedded-Fragmented}}

\begin{figure}[t]
\centering \includegraphics[width=5cm]{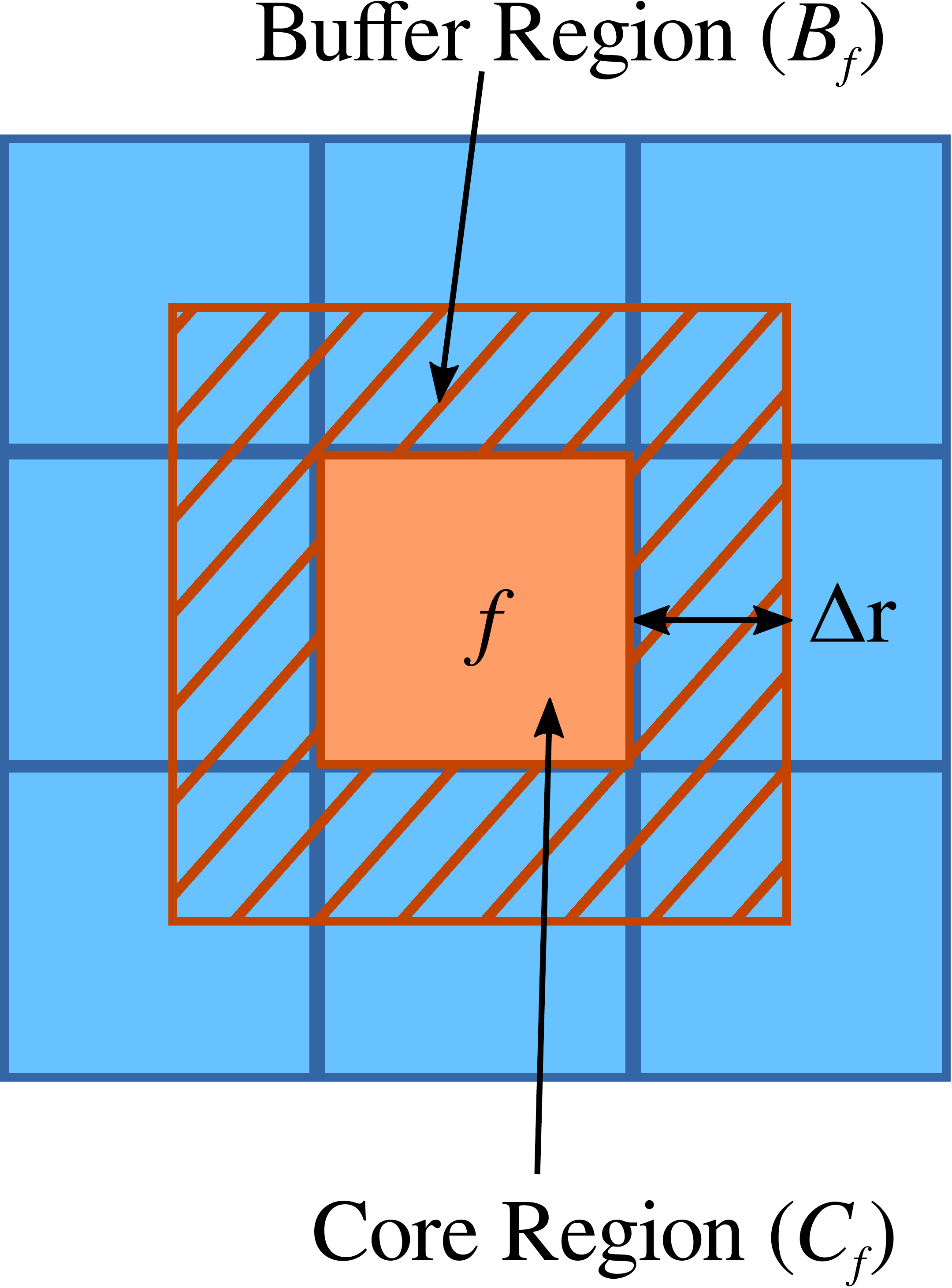} \caption{A sketch of the overlapped fragmentation scheme. The system is first
divided into non-overlapping core fragments (solid orange and blue
regions) of size $C_{f}$ for each fragment $f$. For each core fragment
$f$ (orange core), we define a buffer region (dashed red region)
of size $B_{f}$. The fragment density is then computed for the core
and buffer region together (dressed fragment, $D_{f}$) using periodic
boundary condition for $D_{f}$ and projected onto the core, as described
in the main text.}
\label{fig:overlap-frag} 
\end{figure}

To overcome the limitation of the efsDFT in covalently bonded fragments,
we propose to use overlapping fragments, as sketched in Fig.~\ref{fig:overlap-frag}.
The basic idea behind the proposed overlapped embedded fragmented
stochastic DFT (o-efsDFT) is that each fragment density and density
matrix is calculated with a buffer region, but the projection of the
fragment density is limited to the core region only. This allows for
continuous densities and density matrices across the boundaries between
the core fragments, thereby resolving the issue discussed above and
allowing for a significant reduction of the statistical error, even
for systems with periodic boundary conditions and a delocalized density
matrix.

As sketched in Fig.~\ref{fig:overlap-frag}, the system is divided
into non-overlapped fragments that are composed of unit cells and/or
small supper cells referred to as ``core regions'', labeled $C_{f}$
for fragment $f$. Each core region is then wrapped with a ``buffer''
region ($B_{f}$) and a dressed fragment (labeled $D_{f}=C_{f}\cup B_{f}$)
is defined as the sum of core and buffered regions. For any $\boldsymbol{r}\in C_{f}$,
the dressed fragment density matrix ($\langle\boldsymbol{r}|\hat{\rho}_{f}|\boldsymbol{r}'\rangle$)
is given by: 
\begin{equation}
\langle\boldsymbol{r}|\hat{\rho}_{f}|\boldsymbol{r}'\rangle=\begin{cases}
\sum_{i=1}^{N_{\text{occ}}^{f}}\langle\boldsymbol{r}|\varphi_{i}^{f}\rangle\langle\varphi_{i}^{f}|\boldsymbol{r}'\rangle & \boldsymbol{r}'\in D_{f}\\
0 & \boldsymbol{r}'\not\in D_{f}
\end{cases}\;\;\ldotp\label{eq:trail-dm}
\end{equation}
In the above equation, $\varphi_{i}^{f}(\boldsymbol{r})$ are KS occupied
orbitals for dressed fragment $f$ obtained by a deterministic DFT
in region $D_{f}$, where we impose periodic boundary condition for
each dressed region separately. The closest distance between the boundaries
$C_{f}$ and $D_{f}$ ($\Delta r$, see Fig.~\ref{fig:overlap-frag})
is chosen so that periodic boundary conditions can be imposed on region
$D_{f}$. Note that the above definition of $\langle\boldsymbol{r}|\hat{\rho}_{f}|\boldsymbol{r}'\rangle$
is not necessarily hermitian and idempotent. However, this turns out
to be insignificant, since the density matrix of the entire system,
remains hermitian regardless of how $\langle\boldsymbol{r}|\hat{\rho}_{f}|\boldsymbol{r}'\rangle$
behaves (see more detail below). In the limit were $\Delta r=0$,
o-efsDFT is identical to efsDFT.\citep{doi:10.1063/1.4890651}

With the definition of $\langle\boldsymbol{r}|\hat{\rho}_{f}|\boldsymbol{r}'\rangle$
in Eq.~\eqref{eq:trail-dm}, the o-efsDFT density at position $\boldsymbol{r}\in C_{f}$
is given by:
\begin{align}
\rho(\boldsymbol{r}) & =\langle\boldsymbol{r}|\hat{\rho}_{f}\hat{\rho}_{f}^{\top}|\boldsymbol{r}\rangle+\langle|\xi(\boldsymbol{r})|^{2}\rangle_{\chi}-\left\langle \langle\boldsymbol{r}|\hat{\rho}_{f}|\chi\rangle\langle\chi|\hat{\rho}_{f}^{\top}|\boldsymbol{r}\rangle\right\rangle _{\chi}\nonumber \\
 & =\rho_{f}(\boldsymbol{r})+\langle|\xi(\boldsymbol{r})|^{2}\rangle_{\chi}-\langle|\xi^{f}(\boldsymbol{r})|^{2}\rangle_{\chi}\;\;,\label{eq:rho-overlap-frag}
\end{align}
where $\xi_{f}(\boldsymbol{r})=\sum_{i=1}^{N_{occ}^{f}}\varphi_{i}^{f}(\boldsymbol{r})\langle\varphi_{i}^{f}|\chi\rangle_{D_{f}}$
and $\langle f|g\rangle_{D_{f}}=\int_{D_{f}}\mathrm{d}\boldsymbol{r}f^{\ast}(\boldsymbol{r})g(\boldsymbol{r})$,
i.e. integration over the dressed fragment region. It is easy to verify
that the density in Eq.~\eqref{eq:rho-overlap-frag} satisfies the
following requirements:
\begin{enumerate}
\item In the infinite sampling limit the properties calculated from o-efsDFT
converge to deterministic DFT results.
\item If $C_{f}$ is the system itself, properties calculated from o-efsDFT
are equal to those from the deterministic DFT calculation regardless
of the number of stochastic orbitals.
\item Given any partitioning of the system into core regions, in the limit
where the buffered zone grows such that $D_{f}$ represents the entire
system, properties calculated from o-efsDFT will, again be equivalent
to the deterministic results with any number of stochastic orbitals.
\end{enumerate}
We now turn to discuss the calculations of kinetic energy, the nonlocal
pseudopotential energy, and the nonlocal force using the above formalism
with overlapped fragments. The calculation of the local pseudopotential
energy, the Hartree term, the exchange correlation energies, and the
corresponding forces can be obtained directly using the density in
Eq.~\eqref{eq:rho-overlap-frag}, similar to sDFT~\citep{PhysRevLett.111.106402}
and efsDFT.\citep{doi:10.1063/1.4890651,doi:10.1063/1.4984931} The
kinetic energy is evaluated in real-space by integrating the kinetic
energy density over the core region of each fragment: 
\begin{align}
E_{K} & =\sum_{f}\sum_{i=1}^{N_{occ}^{f}}\langle\varphi_{i}^{f}|\hat{t}|\varphi_{i}^{f}\rangle_{C_{f}}\nonumber \\
+ & \left\langle \langle\xi|\hat{t}|\xi\rangle\right\rangle _{\chi}-\sum_{f}\left\langle \langle\xi_{f}|\hat{t}|\xi_{f}\rangle_{C_{f}}\right\rangle _{\chi}.\label{eq:overlap-frag-ke}
\end{align}
For an infinite sample set ($N_{\chi}\rightarrow\infty$), $\left\langle \langle\chi|\varphi_{i}^{f}\rangle_{D_{f}}\langle\varphi_{j}^{f}|\chi\rangle_{D_{f}}\right\rangle _{\chi}=\langle\varphi_{j}^{f}|\varphi_{i}^{f}\rangle_{D_{f}}=\delta_{ij}$.
Therefore, the first and third terms in Eq.~\eqref{eq:overlap-frag-ke}
cancel each other. Similarly, the non-local pseudopotential energy
($E_{\text{nl}}$) can be calculated as follows:
\begin{align}
E_{\text{nl}} & =\sum_{f}\sum_{\text{atom}\in C_{f}}\sum_{i=1}^{N_{occ}^{f}}\langle\varphi_{i}^{f}|\hat{v}_{\text{nl}}^{\text{atom}}|\varphi_{i}^{f}\rangle_{D_{f}}+\left\langle \langle\xi|\hat{v}_{nl}|\xi\rangle\right\rangle _{\chi}\nonumber \\
- & \sum_{f}\sum_{\text{atom}\in C_{f}}\left\langle \langle\xi_{f}|\hat{v}_{\text{nl}}^{\text{atom}}|\xi_{f}\rangle_{D_{f}}\right\rangle _{\chi},\label{eq:overlap-frag-enl}
\end{align}
where $\hat{v}_{\text{nl}}^{\text{atom}}$ is the non-local pseudopotential
operator for each atom and the sum $\sum_{\text{atom}\in C_{f}}$
refers to summation over all atoms that are in region $C_{f}$. The
non-local pseudopotential nuclear force for each atom with a position
$R_{\text{atom}}\in C_{f}$, is given by:
\begin{align}
\boldsymbol{F}_{\text{nl}}^{\text{atom}} & =\sum_{i=1}^{N_{occ}^{f}}\left\langle \varphi_{i}^{f}\left|\frac{\partial\hat{v}_{nl}^{\text{atom}}}{\partial\boldsymbol{R}_{\text{atom}}}\right|\varphi_{i}^{f}\right\rangle _{D_{f}}\nonumber \\
+ & \left\langle \left\langle \xi\left|\frac{\partial\hat{v}_{nl}^{\text{atom}}}{\partial\boldsymbol{R}_{\text{atom}}}\right|\xi\right\rangle \right\rangle _{\chi}-\left\langle \left\langle \xi_{f}\left|\frac{\partial\hat{v}_{nl}^{\text{atom}}}{\partial\boldsymbol{R}_{\text{atom}}}\right|\xi_{f}\right\rangle _{D_{f}}\right\rangle _{\chi}.\label{eq:overlap-frag-fnl}
\end{align}
It is straightforward to show that Eqs.~\eqref{eq:overlap-frag-enl}
and \eqref{eq:overlap-frag-fnl} satisfy all three requirements above. 

\section{Implementation}

\begin{table*}
\begin{tabular}{cccllll}
\hline 
System  & Fragment  & $N_{\chi}$  & $E_{K}/N_{e}$  & $E_{\text{nl}}/N_{e}$  & $E_{\text{loc}}/N_{e}$  & $E_{\text{tot}}/N_{e}$ \tabularnewline
\hline 
Si$_{512}$  & Deterministic &  & 10.3577  & 6.1274  & -8.6267  & -26.9954 \tabularnewline
 & Si$_{8}$  & 800  & 10.3596(88)  & 6.1269(29)  & -8.6215(41)  & -26.9881(77) \tabularnewline
 & Si$_{64}$  & 80  & 10.3523(21)  & 6.1277(59)  & -8.6223(61)  & -26.9955(12) \tabularnewline
 & Si$_{216}$  & 80  & 10.3560(13)  & 6.1284(15)  & -8.6265(27)  & -26.9959(5) \tabularnewline
Si$_{1728}$  & Si$_{64}$  & 80  & 10.3507(14)  & 6.1287(25)  & -8.6207(31)  & -26.9953(11) \tabularnewline
 & Si$_{216}$  & 80  & 10.3527(5)  & 6.1286(7)  & -8.6227(6)  & -26.9964(2) \tabularnewline
Si$_{4096}$  & Si$_{64}$  & 80  & 10.3526(23)  & 6.1286(5)  & -8.6219(13)  & -26.9944(12) \tabularnewline
 & Si$_{216}$  & 80  & 10.3539(7)  & 6.1296(7)  & -8.6243(11)  & -26.9957(5) \tabularnewline
\hline 
\end{tabular}\caption{\label{table:energy} Si$_{512}$, Si$_{1728}$, Si$_{4096}$ were
calculated by o-efsDFT with Si$_{64}$/Si$_{216}$ as fragments. Kinetic
energy per electron, non-local pseudo-potential energy per electron,
and total energy per electron (in eV) are presented in the table.
The standard deviation in the last digits of energies are listed in
parenthesis. Deterministic DFT calculation of Si$_{512}$ is also
presented.}
\end{table*}

The o-efsDFT approach can be implemented for any planewave or real-space
based approach. The necessary steps to complete the self-consistent
iteration to converge the density, energy and forces, can be summarized
as follows (for a planewave based approach):
\begin{enumerate}
\item Perform a deterministic DFT calculation in region $D_{f}$ for each
fragment $f$ using a real-space grid spacing that equals that the
grid spacing of the full system. 
\item Using the fragment density as an initial guess, generate the corresponding
Kohn-Sham potential.
\item Generate a set of stochastic orbitals $|\chi_{1}\rangle\cdots|\chi_{N_{\chi}}\rangle$
on a real-space grid and then transform to reciprocal space (it requires
$\approx N_{g}/2$ grid points in reciprocal space to store an $N_{g}$
real-space orbital due to the isotropic kinetic energy cutoff).
\item For all stochastic orbital $|\chi\rangle$, expand the action of the
density matrix in terms of Chebyshev polynomials and tune the chemical
potential to satisfy:
\begin{align}
N_{e}(\mu) & +\sum_{f}\int_{C_{f}}\mathrm{d}\boldsymbol{r}\rho^{f}(\boldsymbol{r})\nonumber \\
- & \sum_{f}\int_{C_{f}}\mathrm{d}\boldsymbol{r}\langle|\xi_{f}(\boldsymbol{r})|^{2}\rangle_{\chi}=N_{e}\,\,,\label{eq:chem-pot}
\end{align}
where as before, $N_{e}(\mu)=\left\langle \left\langle \chi\left|\theta_{\beta}(\mu-\hat{h}_{\text{KS}})\right|\chi\right\rangle \right\rangle _{\chi}$,
$\rho_{f}(\boldsymbol{r})$ is define in Eq.~\eqref{eq:rho-overlap-frag},
$\xi_{f}\left(\boldsymbol{r}\right)=\sum_{i=1}^{N_{occ}^{f}}\varphi_{i}^{f}(\boldsymbol{r})\langle\varphi_{i}^{f}|\chi\rangle_{D_{f}}$,
and $N_{e}$ is the number of electrons in the system. 
\item Use the chosen $\mu$ from the previous step to generate a set of
stochastic project orbitals, $|\xi\rangle=\sqrt{\theta_{\beta}(\mu-\hat{h}_{\text{KS}})}|\chi\rangle$,
using a Chebyshev or Newton interpolation polynomials to act with
$\sqrt{\theta_{\beta}(\mu-\hat{h}_{\text{KS}})}$ on $|\chi\rangle$.
\item Calculate the density $\rho(\boldsymbol{r})$ using Eq.~\eqref{eq:rho-overlap-frag}
and the corresponding energy with Eqs.~\eqref{eq:overlap-frag-ke}
and \eqref{eq:overlap-frag-enl}. The other energy terms can be obtained
directly from the density, similarly to the deterministic DFT.
\item Use the density as input for the next iteration and go to step (d).
Repeat until converges is achieved.
\end{enumerate}
Throughout the self-consistent iterations the same random number seed
was used to generate the stochastic orbitals. This is necessary to
converge the self-consistent procedure to a tolerance similar to deterministic
DFT. This also reduces the computational effort associated with projecting
the stochastic orbitals onto the fragments, so that the first and
last terms in Eqs.~\eqref{eq:rho-overlap-frag} -- \eqref{eq:overlap-frag-fnl}
were calculated prior to the self-consistent loop.

\section{Results and Discussion\label{sec:Results-and-Discussion}}

To assess the accuracy of the o-efsDFT method and its limitations,
we performed $\Gamma$ point calculations on a set of silicon crystals
with varying super-cell size. Bulk silicon is rather challenging for
linear scaling DFT due to its particularly small LDA band gap.\citep{doi:10.1063/1.2796168}
A grid spacing of 0.41$a_{0}$ corresponding to a planewave cutoff
of $60$~Ryd was used to converge the energy and force to within
an acceptable tolerance. To reduce the energy range of the KS-Hamiltonian,
we used an ``early truncation'' scheme for the kinetic energy operator
in which the magnitude of the reciprocal space vector $\boldsymbol{k}$
(used in evaluating the kinetic energy operator acting on a wave function)
was replaced by $\min\{\|\boldsymbol{k}\|,k_{\text{ke}}\}$, where
$k_{\text{ke}}=\sqrt{2E_{\text{cut}}^{\text{ke}}}$, with $E_{\text{cut}}^{\text{ke}}=20$~Ryd
for all $3$ systems studied here. The Troullier-Martins norm conserving
pseudopotentials~\citep{PhysRevB.43.1993} were used together with
the Kleinman-Bylander separable form.\citep{PhysRevLett.48.1425}
The pseudopotentials were evaluated in real-space to reduce the computational
scaling.\citep{PhysRevB.44.13063} In the Fermi function $\beta$
was set to $\approx600$ inverse Hartree, which is sufficient to converge
the ground state properties to within an error much smaller than the
statistical error. 

\begin{figure}[b]
\centering \includegraphics[width=8cm]{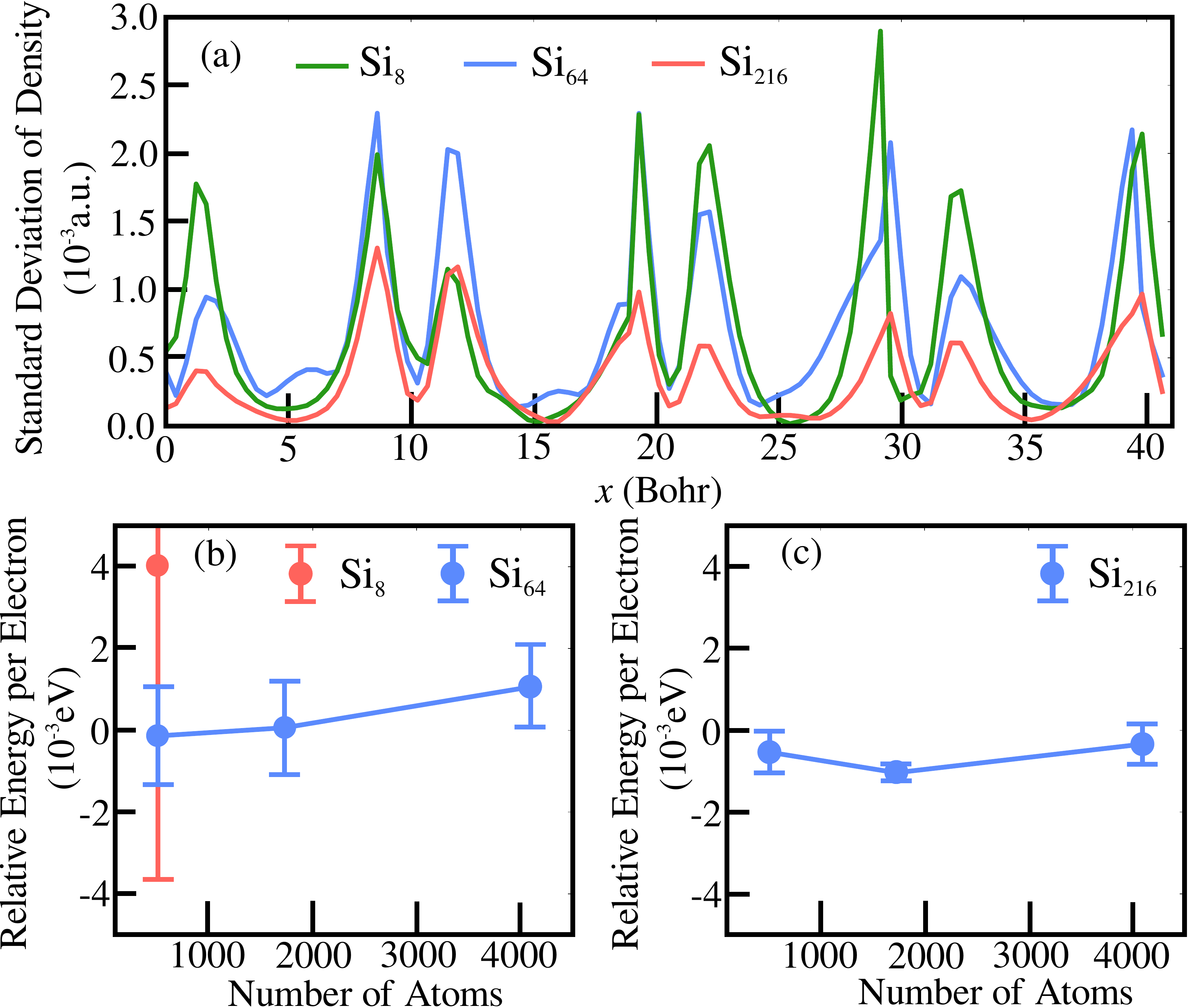} \caption{\label{fig:rho-ene-conv} Panel (a): The standard deviation of the
density along the $x$-axis for $y=z=0$ for $\text{Si}_{512}$ with
three different dressed fragment sizes (all for $\text{Si}_{8}$ core
fragment), as specified in the legend. Panels (b) and (c): The energy
per electron (relative to that of a deterministic DFT calculation
for $\text{Si}_{512}$) as a function of the system size for dressed
fragment of $\text{Si}_{64}$(panel (b)) and $\text{Si}_{216}$ (panel
(c)) with $N_{\chi}=80$. In panel (b) we also show the result with
$\text{Si}_{8}$ dressed fragment (namely, no overlap between the
fragments) with $N_{\chi}=800$, where we also observe a large bias~\citep{Roisubmit}
and find that the energy is larger by $4$~eV per atom compared to
the deterministic value.}
\end{figure}

We studied bulk silicon with $3$ supper cells of size $4\times4\times4$
unit cells ($N_{\text{atom}}=512$), $6\times6\times6$ unit cells
($N_{\text{atom}}=1728$), and $8\times8\times8$ unit cells ($N_{\text{atom}}=4096$).
Core fragments of $\text{Si}_{8}$were used to partition all systems
while three different buffer regions were adopted for forming the
dressed fragments with $8$, $64$, and $216$ Si atoms. $N_{\chi}=80$
stochastic orbitals were used in each calculation but for the non-overlapped
Si$_{8}$ fragments, we used $N_{\chi}=800$ stochastic orbitals.
It should be noted that calculations with $N_{\chi}=80$ stochastic
orbitals failed to converge without using overlapped fragments. The
self-consistent iteration were converged using DIIS~\citep{Pulay1980,Pulay1982,Kresse1996}
with a convergence criteria set to $10^{-8}$ Hartree per electron. 

Table~\ref{table:energy} summarizes the results for the kinetic,
nonlocal, local, and total energies per electron for all three system
sizes studied and for all dressed fragments used. The external local
potential energy per electron, $E_{\text{loc}}/N_{e}$, depends linearly
on the density and provides a reliable measure of the accuracy of
the stochastic density. As can be seen in Table~\ref{table:energy},
for Si$_{512}$, $E_{\text{loc}}/N_{e}$ with dressed fragments of
sizes Si$_{64}$ and Si$_{216}$ the external potential energy per
electron agrees well with the deterministic DFT result, where the
differences between the two are well within the error bars of $5$
independent o-efsDFT runs. Reductions in the standard deviation of
$E_{\text{loc}}/N_{e}$ with respect to the dressed fragment size
was observed in all $3$ system sizes. 

\begin{figure}[t]
\centering \includegraphics[width=7cm]{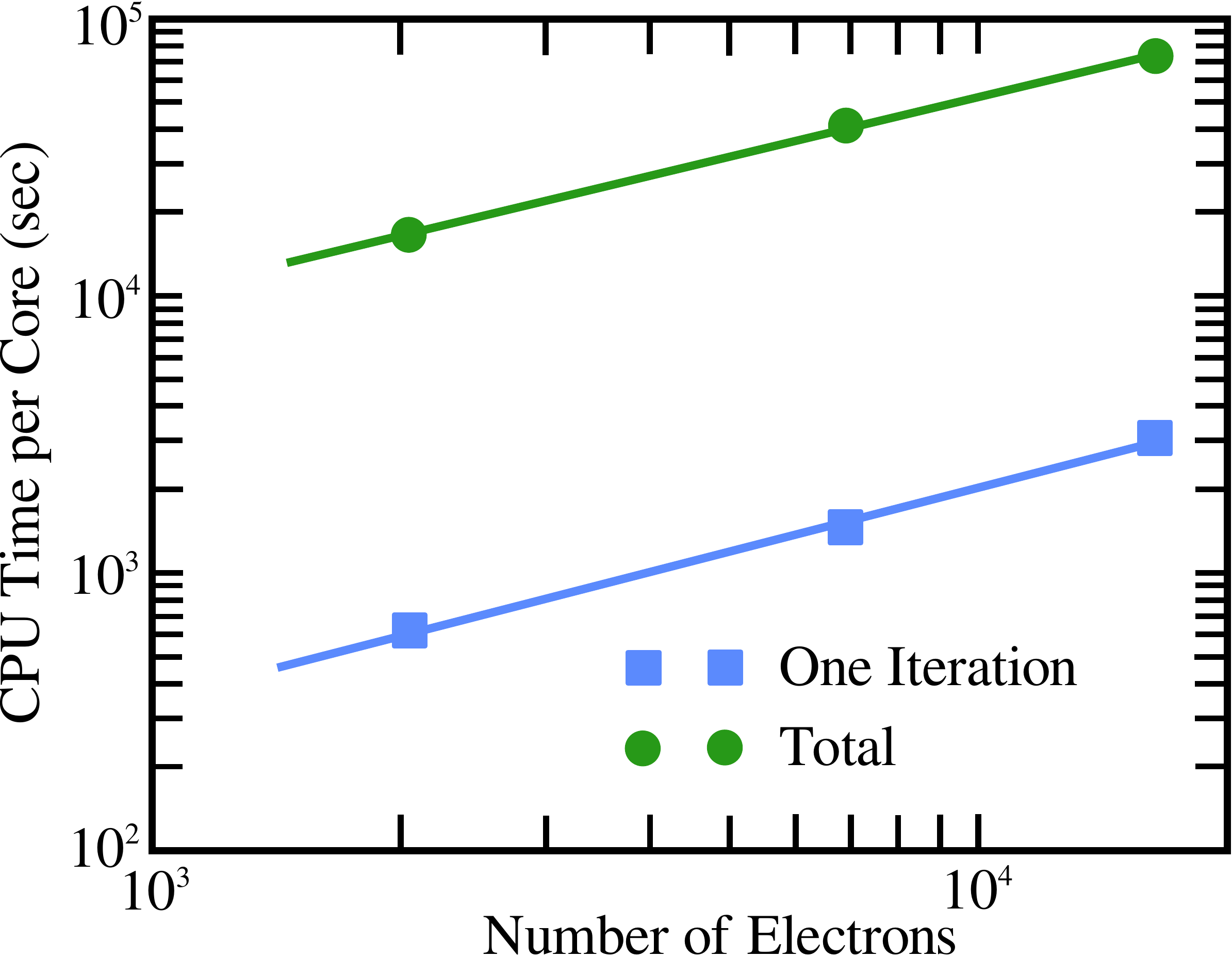} \caption{\label{fig:scaling} A log-log plot of the CPU time per core versus
the number of electrons. The green and blue symbols are for the total
wall time of all self-consistent iterations and for a single self-consistent
iteration, respectively. The straight lines are fits to a power law
with exponents of $N_{e}^{0.71}$ and $N_{e}^{0.77}$, respectively. }
\end{figure}

This is also illustrated in panel (a) of Fig.~\ref{fig:rho-ene-conv},
where we plot the density along the $x$-axis (at fixed $y$ and $z$)
for Si$_{512}$ for all three dressed fragment sizes. The non-overlapping
fragments required $10$ times more stochastic orbitals ($N_{\chi}=800$)
in order to reduce the noise in the density to similar levels as the
overlapping Si$_{64}$ dressed fragments. For Si$_{216}$ dressed
fragments, the noise in the density dropped by a factor of $\approx2$,
suggesting that only $N_{\chi}=20$ stochastic orbitals are needed
to converge the density to the same level of noise as that of Si$_{64}$
dressed fragments. 

The decrease in the noise with increasing buffer sizes (increase $B_{f}$
region) is not limited to the density itself or to local observables
that depends directly on the density, like $E_{\text{loc}}/N_{e}$.
The kinetic energy per electron ($E_{K}/N_{e}$) and non-local pseudopotential
energy per electron ($E_{\text{nl}}/N_{e}$) also show a significant
reduction in their variance with increasing $B_{f}$, as can be seen
in Table~\ref{table:energy}. Comparing to the deterministic results
for Si$_{512}$ we find that o-efsDFT provides energies per electron
that are within $10^{-3}$~eV from the deterministic values and that
the agreement improves with the size of the buffer regions. This is
also illustrated in panels (b) and (c) of Fig.~\ref{fig:rho-ene-conv},
where we plot the total energy per electron (relative to a reference
energy of the deterministic result for Si$_{512}$) for the $3$ system
sizes and for $3$ different buffer regions (Si$_{8}$,Si$_{64}$,
and Si$_{216}$) studied in this work. As before, we use $N_{\chi}=80$
stochastic orbitals except for $\text{Si}_{8}$ buffer region, where
$N_{\chi}=800$. The total energy per electron contains much larger
noise when the fragments do not overlap, even when the number of stochastic
orbitals was $10$ larger. Replacing the Si$_{8}$ fragment with a
dressed Si$_{64}$ fragment significantly reduces the noise in the
total energy per electron and further reduction in the noise was achieved
by enlarging the buffer region, i.e. using Si$_{216}$ as dressed
fragment. 

\begin{figure}[t]
\centering \includegraphics[width=8cm]{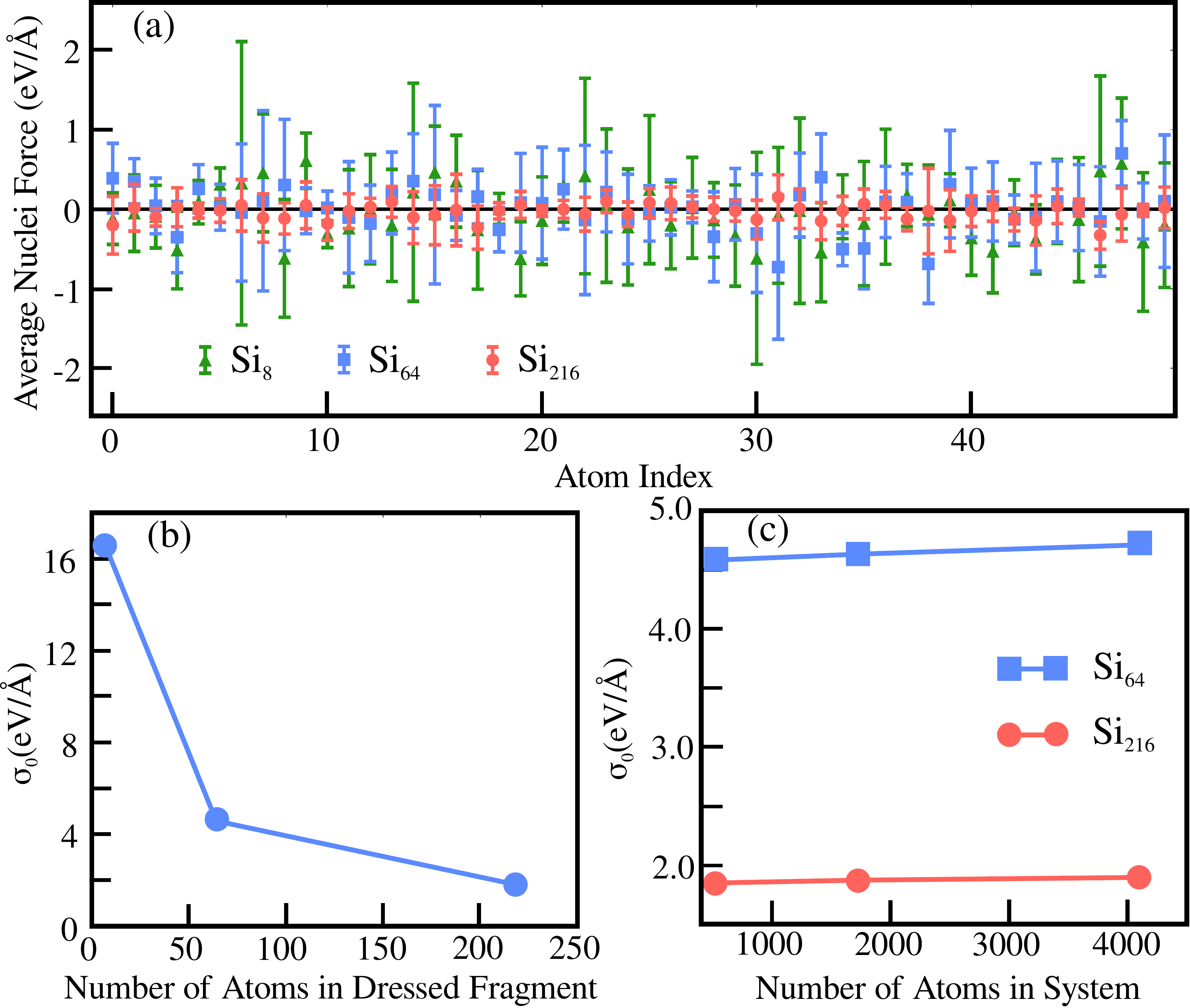} \caption{\label{fig:force-all} Panel (a): Nuclei forces along $x$-axis for
$50$ representative atoms in Si$_{512}$ with Si$_{64}$ (blue) and
Si$_{216}$ (red) dressed fragments and for $N_{\chi}=80$. The dots
and error bars are averages and standard deviations of the forces
from $5$ independent runs. Panel (b) and (c) show the value of $\sigma_{0}$
(see the main text for definition) as a function of the buffer size
and the number of electrons, respectively.}
\end{figure}

Within the accuracy of the current calculations, we find that the
statistical error in the total energy per electron is similar for
all $3$ system sizes studied here (for a fixed buffer size), implying
that self-averaging is not significant.\citep{PhysRevLett.111.106402}
This suggests that o-efsDFT scales nearly linearly with the system
size, as indeed is observed for the scaling of the energy per electron,
shown in Fig.~\ref{fig:scaling}. We focused on the computational
cost of the self-consistent iterations which dominates the overall
CPU time. Since parallelism over stochastic orbitals is straightforward
and the communication time is negligible, we report the CPU time per
stochastic orbital (each orbital was distributed on a different node).
Our results show that the time used in each self-consistent scales
as $O\left(N_{e}^{0.77}\right)$. The net CPU time per stochastic
orbital over all self-consistent iterations scales as $O\left(N_{e}^{0.71}\right)$,
which suggests that the number of self-consistent iterations does
not significantly increases with increasing system size. We note that
the practical scaling is somewhat better than linear ($O\left(N_{e}\right)$)
mainly due to improved multithreading timings for larger grids. This
however, may depend on the computational architecture used and on
the range of systems studied. Generally, we expect the approach to
scale as $O\left(N_{e}\right)$ for large systems.

\begin{figure*}[t]
\centering \includegraphics[width=1\textwidth]{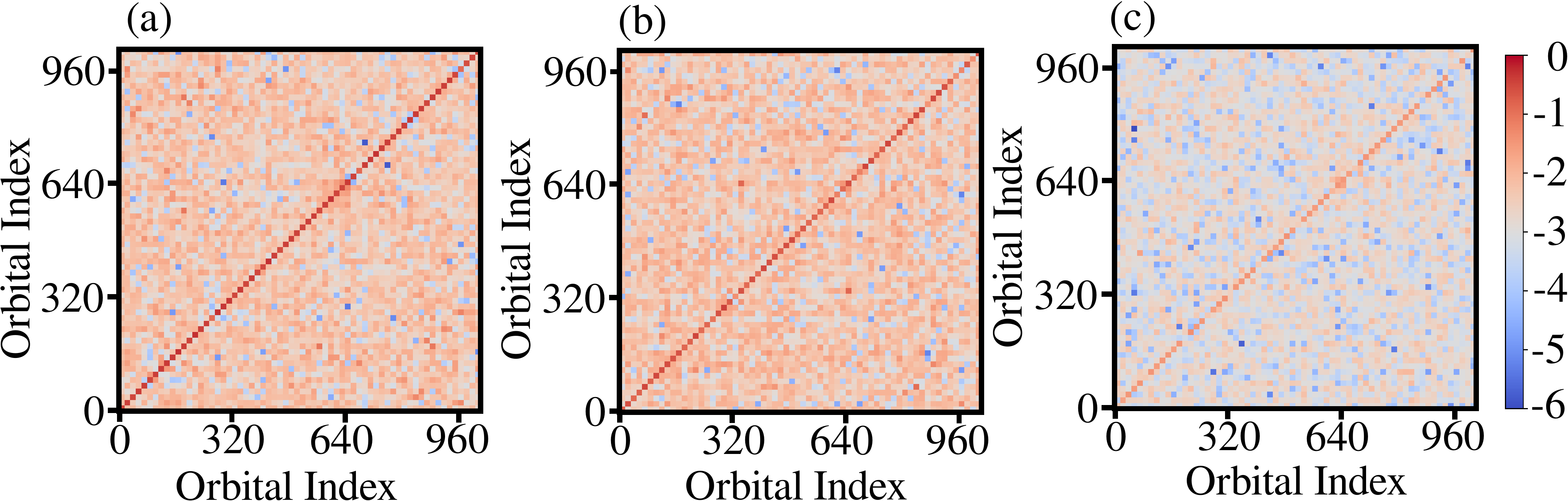} \caption{\label{fig:dm-qual} Fragment density matrices difference $\Delta\hat{\rho}=\sum_{_{f}}\hat{\rho}_{f}-\hat{\rho}$
evaluated for Si$_{512}$ with (a) non-overlapped Si$_{8}$ fragment,
(b) non-overlapped Si$_{64}$ fragment, and (c) overlapped Si$_{64}$
fragments. Fragment density matrices are represented in the subspace
spanned by occupied orbitals of Si$_{512}$ In this representation,
the system density matrix $\hat{\rho}$ becomes the identity matrix.
To span several orders of magnitude in the values $\Delta\hat{\rho}$
we use a logarithmic color scale. }
 
\end{figure*}

In addition to computing the individual contribution to the total
energy per electron, o-efsDFT also offers an efficient and accurate
approach to compute the forces on each atom. Panel (a) of Fig.~\ref{fig:force-all}
shows the average and standard deviation of nuclear forces from $5$
independent o-efsDFT runs for $50$ representative atoms in Si$_{512}$.
The forces are computed for the equilibrium structure and thus should
fluctuate about $0$. Forces from o-efsDFT with non-overlapped fragment
and $N_{\chi}=800$ stochastic orbitals fluctuate with comparable
standard deviations as those from o-efsDFT with Si$_{64}$ dressed
fragments and $N_{\chi}=80$ stochastic orbitals, yet the computational
effort is nearly $10$ times smaller in the latter. Smaller force
fluctuations were observed with growing buffer region, similar to
the reduction in the statistical noise in the total energy per electron.

A summary of the statistical error in the forces as a function of
the size if the dressed fragment and as a function of the system size
are shown in panels (b) and (c) of Fig.~\ref{fig:force-all}, respectively.
Within the central limit theorem, we assume that the standard deviation
of nuclei force decays as $\sigma_{0}/\sqrt{N_{\chi}}$ where $\sigma_{0}$
is the standard deviation with only $1$ stochastic orbital. In panel
(b) we plot $\sigma_{0}$ as a function of the dressed fragment size,
where a significant reduction is clearly observed. Panel (c) shows
that $\sigma_{0}$ does not change significantly with system size,
suggesting that the number of stochastic orbitals does not increase
with the system size for a given statistical error, suggesting that
the scaling for the force follows that of the energy per electron
shown in Fig.~\ref{fig:scaling}.

\section{Overlapped versus Non-overlapped Fragments}

To further understand the o-efsDFT results presented in the previous
section we examine the role played by the overlap of fragments. In
the supplementary information we derive an expression for the variance
of the density in terms of the density itself and the density matrix:
\begin{align}
\text{Var}\left\{ \rho\left(r\right)\right\}  & \leq\left(2\max_{\boldsymbol{r}'}\langle\boldsymbol{r}'|2\hat{\rho}+\Delta\hat{\rho}^{\top}|\boldsymbol{r}\rangle^{2}+4\rho(\boldsymbol{r})\right)\nonumber \\
\times & \int\mathrm{d}\boldsymbol{r}'\langle\boldsymbol{r}|\Delta\hat{\rho}|\boldsymbol{r}'\rangle^{2}\label{eq:var-5-1}
\end{align}
where $\Delta\hat{\rho}=\sum_{f}\hat{\rho}_{f}-\hat{\rho}$ and $\hat{\rho}_{f}$
and $\rho$ are defined in Eq.~\eqref{eq:trail-dm} and above Eq.~\eqref{eq:rho},
respectively. It is clear that the noise in the density is not just
correlated with the density, but also with the density matrix. Thus,
fragmentation schemes to reduce the noise in $\rho\left(\boldsymbol{r}\right)$
must provide a reasonable approximation for the density matrix across
different fragments. This is not the case for the non-overlap fragments,
where off-diagonal matrix elements of the full reference density matrix
($\sum_{f}\hat{\rho}_{f}$) for different fragments vanish, leading
to a significant level of noise. In o-efsDFT with non-vanishing buffer
zones, the reference density matrix is allowed to decay at least within
distance $\Delta r$ before it is truncated to $0$ (see Fig.~\eqref{fig:overlap-frag}).
This improves the reference density matrix and the resulting noise
is significantly reduced as the size of the buffer zone increases. 

In Fig.~\ref{fig:dm-qual} we show the reference density matrix for
three different fragmentation schemes. Panels (a) and (b) correspond
to non-overlapped fragments with $\text{Si}_{8}$ and $\text{Si}_{64}$
fragments, respectively and panel (c) shows the reference density
matrix for $\text{Si}_{8}$ with a dressed fragment of $\text{Si}_{64}$.
In the eigenfunction representation, the density matrix should equal
the identity matrix. This is clearly not the case for panel (a) and
(b), but is rather close to the unit matrix for panel (c). This is
rather surprising since even with fragments that equal the size of
the the dressed fragment, the density matrix does not improve. Obviously,
the discontinuity in the density matrix at the interfaces between
fragments does not depends on the size of the fragment for the non-overlapped
case, leading to significant noise in the density even for large fragments.
While it only takes a small fragment, $\text{Si}_{8}$, with a buffer
zone of $\text{Si}_{64}$ to significantly reduce the noise.

\section{Conclusion}

Motivated by the need to reduce significantly the noise level in linear
scaling sDFT, we have proposed a new fragmentation scheme that circumvents
some of limitations of our original embedded-fragmented stochastic
DFT approach, mainly for covalently bonded systems. While previous
attempts to develop a fragmentation scheme were based solely on providing
a good reference system for the density alone, here, we showed that
this is not a sufficient criteria. Based on detailed analysis of the
variance in the density, we developed a new overlapped fragmentation
scheme, which provides an excellent reference for both the \emph{density
matrix} and the \emph{density}. By dividing the system into non-overlapped
core fragments and overlapped dressed fragments, we demonstrated that
the existence of a buffer zone is crucial in the reduction of the
stochastic noise in the energy per electron and nuclear forces. Moreover,
we showed that the stochastic noise does not increase with respect
to system size when using a fixed fragment sizes nor does the number
of self-consistent iterations, leading to a practical scaling that
is slightly better than the expected linear scaling. While all application
in this study were limited to bulk silicon with periodic boundary
condition, we wish to iterate that the approach is also valid for
open boundary conditions, were we expect it to work equally well.
\begin{acknowledgments}
RB gratefully acknowledges support from the Israel Science Foundation,
grant No. 189/14. DN and ER are grateful for support by the Center
for Computational Study of Excited State Phenomena in Energy Materials
(C2SEPEM) at the Lawrence Berkeley National Laboratory, which is funded
by the U.S. Department of Energy, Office of Science, Basic energy
Sciences, Materials Sciences and Engineering Division under contract
No. DEAC02-05CH11231 as part of the Computational Materials Sciences
Program. Resources of the National Energy Research Scientific Computing
Center (NERSC), a U.S. Department of Energy Office of Science User
Facility operated under Contract No. DE-AC02-05CH11231 are greatly
acknowledged. 
\end{acknowledgments}

%\bibliographystyle{aipnum4-1}
%\bibliography{o-efsDFT}
%

\end{document}